\documentclass[10pt,letterpaper]{article}
\usepackage{opex3}
\usepackage[utf8x]{inputenc}
\usepackage[T1]{fontenc}

\usepackage{graphicx}
\usepackage{xcolor}
\usepackage[binary-units]{siunitx}
\usepackage{multirow}
\usepackage[bookmarks=true]{hyperref}
\usepackage{amsmath}
\usepackage{cite}

\begin{document}

\title{Free-space quantum key distribution to a moving receiver}
\author{Jean-Philippe Bourgoin,$^{1,2,4}$ Brendon~L. Higgins,$^{1,2}$ Nikolay Gigov,$^{1,2}$ Catherine Holloway,$^{1,2}$ Christopher~J. Pugh,$^{1,2}$ Sarah Kaiser,$^{1,2}$ Miles Cranmer$^{1,2}$ and Thomas  Jennewein$^{1,2,3,*}$}
\address{$^{1}$Institute for Quantum Computing, University of Waterloo, Waterloo, ON N2L 3G1, Canada\\
$^{2}$Department of Physics and Astronomy, University of Waterloo, Waterloo, ON N2L 3G1, Canada\\
$^{3}$Quantum Information Science Program, Canadian Institute for Advanced Research, Toronto, ON, Canada\\
$^{4}$jbourgoin@uwaterloo.ca}
\email{$^{*}$thomas.jennewein@uwaterloo.ca}

\begin{abstract}
Technological realities limit terrestrial quantum key distribution (QKD) to single-link distances of a few hundred kilometers. One promising avenue for global-scale quantum communication networks is to use low-Earth-orbit satellites. Here we report the first demonstration of QKD from a stationary transmitter to a receiver platform traveling at an angular speed equivalent to a \SI{600}{\km} altitude satellite, located on a moving truck. We overcome the challenges of actively correcting beam pointing, photon polarization and time-of-flight. Our system generates an asymptotic secure key at \SI{40}{bits/\s}.
\end{abstract}
\ocis{(270.5568) Quantum cryptography; (060.2605) Free-space optical communication.}


\section{Introduction} \label{sec.Intro}

Contemporary communications security relies on assumptions about an adversary's computational ability, a property that is uncontrolled and can rapidly change. In contrast, quantum key distribution (QKD) provides a mechanism to establish future-proof secure communications between two parties by exploiting fundamental quantum mechanics to generate a secret, random encryption key common to both parties~\cite{SBCDLP09}. While commercial application of this technology is growing, with current technology direct QKD links on the ground cannot reach distances beyond a few hundred kilometers due to optical losses~\cite{UTSWSLBJPTOFMRSBWZ07, SWVTGZGTT09, LCWCWCWLLYPCCP10, ZHHJIJ12,KLHGLNSTZ15}. Quantum repeaters~\cite{BBCJPW93,BDCZ98} promise to be an essential component of future long-distance QKD networks, but such devices are not ready for operational integration~\cite{SAABDGHJKMNPRDRSSSTWWWWY10, SSRG11}. Alternatively, an orbiting satellite could serve as an untrusted node~\cite{HBKLMNP00, ZCSCWL14}---linking two ground stations simultaneously and facilitating key distribution without acquiring the key itself---or as trusted node~\cite{BHLMNP00, NHMPW02, RTGK02,  UJKPCMAVSABBBCCGLHLLMPRRRSSTTTOPVWWWZZ08, VJTABUPLBZB08, EAWANS11, HN11, MYMBHJ11,YCLPWYZYCPP13}---exchanging individual keys with each ground station and broadcasting the combination of two keys to allow two ground stations to establish a shared key. Although the satellite is privy to the keys, by requiring only one link at a time the trusted node satellite benefits from a simpler design and, with a suitable orbit, allows key distribution between two parties located anywhere on Earth.

A satellite QKD link will require optical pointing mechanisms compatible with the requirements of a quantum link---i.e., capable of preserving the quantum information while maximizing the average transmission (but not necessarily the signal consistency required for classical optical communication). Recent demonstrations of QKD using moving \emph{transmitters}~\cite{MNFHRW12, WYLZSHWYTZLLHHRPYJCPP12} imitate the case of a transmitting satellite platform, albeit without a traveling platform reaching the angular speed of a low-Earth-orbit (LEO) satellite. In contrast, a satellite-based quantum \emph{receiver} is somewhat less complex to develop, and adds flexibility to utilize multiple different source technologies, including attenuated laser pulses and entangled photons~\cite{BMHHEHKHDGLJ13}. To show the viability of such an approach, we demonstrate an optical QKD link to a moving receiver platform traveling at equivalent angular speed of a satellite. Our demonstration improves upon previous moving-transmitter-QKD demonstrations by employing a coincidence algorithm incorporating time-of-flight correction, implementing automated correction of polarization drifts (including those due to the movement of the tracking system) incurred in the fiber from the source to the transmitter, and fully conducting the post-processing steps of decoy-state QKD necessary to extract a secure key.

\begin{figure}[tbp]
  \centering
  \includegraphics[width=\linewidth]{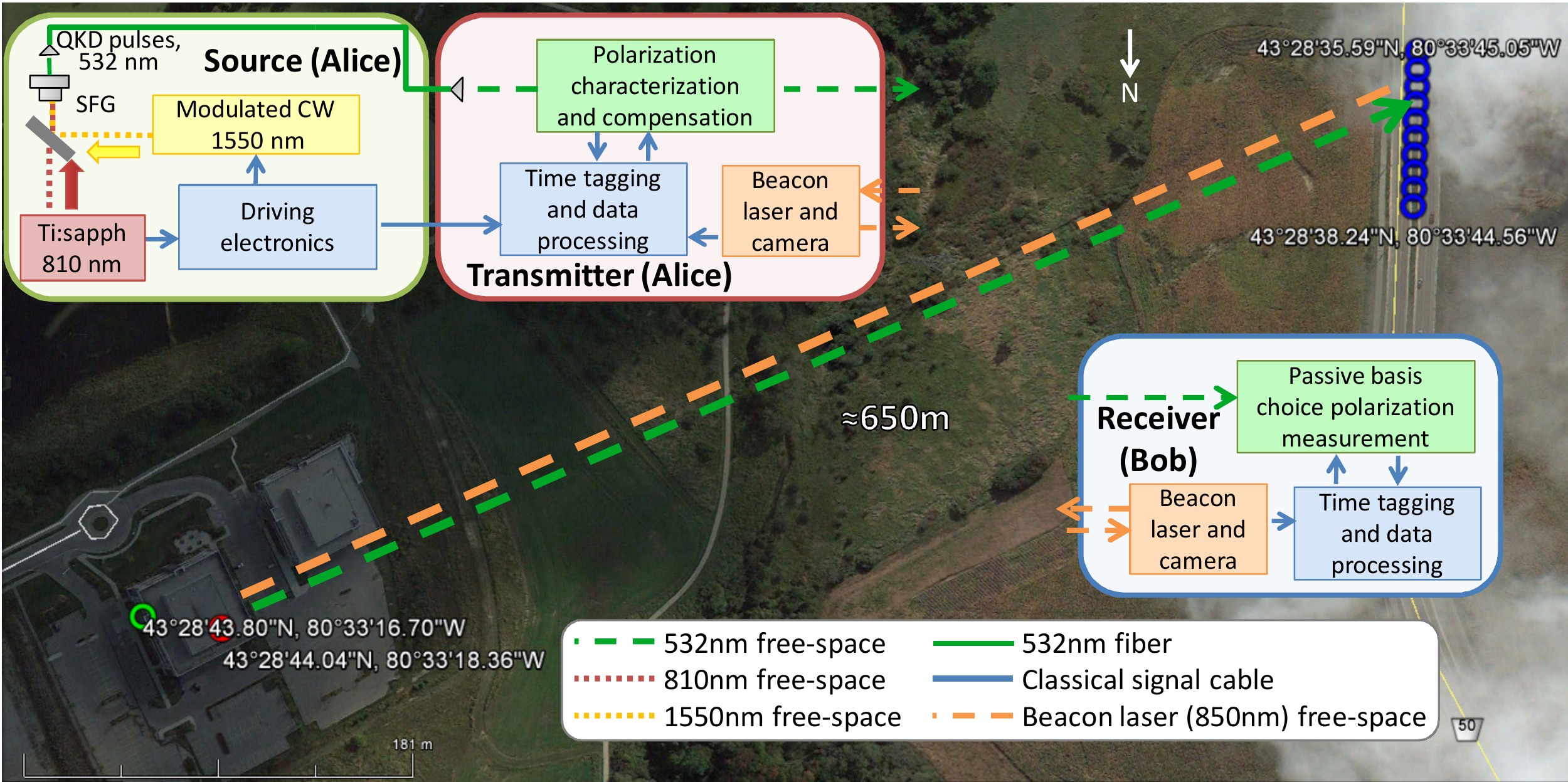}
  \caption{Schematic overview of the moving receiver experimental setup with map showing the location of Alice, consisting of the source (green circle) and transmitter (red circle), and the section of the road that Bob, located on the truck, traveled (blue circles, one per second) during the moving receiver tests. Signals from the source (located in the laboratory on the ground floor of the building) are sent to the transmitter using an optical fiber. An active laser-beacon tracking and pointing system is used to maintain the free-space link while the truck is traveling, and a wireless local area network (WLAN) is used for classical communications. The distance from the transmitter to the truck is $\SI{{\approx}650}{\m}$ and the length of the road traveled during the test was $\SI{{\approx}80}{\m}$. Map data: Google, DigitalGlobe.}
  \label{fig.overview_diagram}
\end{figure}

Our QKD receiver (Bob) is located in the cargo area of a pickup truck driven along a road approximately \SI{650}{\m} from the transmitter (Alice)---see Fig.~\ref{fig.overview_diagram}. The experiment was performed at night to reduce background light contributions and avoid daytime traffic. To maintain the link we utilize \SI{850}{\nm} wavelength beacon lasers and CMOS cameras, in conjunction with a tracking algorithm operating at \SI{24}{\Hz}, at both transmitter and receiver. Over the link we transmit intensity- and polarization-modulated pulses, implementing BB84 QKD with decoy states~\cite{H03}. Simultaneously, the truck maintains an angular speed exceeding the maximum angular speed of a typical \SI{600}{\km} LEO satellite platform~\cite{BMHHEHKHDGLJ13}, thereby supporting the feasibility of performing QKD from a ground station to an orbiting satellite.

\section{Apparatus}

QKD states are generated by a weak coherent pulse (WCP) source located in a ground-floor laboratory in the Research Advancement Center 1 building on the University of Waterloo campus. Our WCP source produces photon pulses at \SI{532}{\nm} wavelength through sum-frequency generation (SFG), combining an \SI{80}{\MHz} pulsed \SI{810}{\nm} Ti:sapphire laser with a continuous wave (CW) \SI{1550}{\nm} laser that is intensity and polarization modulated using fiber-based electro-optical intensity and phase modulators~\cite{YMBHGMHJ13}. The resulting \SI{532}{\nm} pulses possess the same pulse rate and pulse width (${\approx}\SI{0.5}{\ps}$) as the mode-locked Ti:sapphire laser while having the same polarization as the CW laser. The benefit of using SFG to create \SI{532}{\nm} pulses is that the short coherence length of the \SI{1550}{\nm} laser removes the phase correlation between the pulses (an inherent property of a mode-locked laser) while allowing the use of fast and stable fiber-based electro-optical modulators designed for the telecom-band.

\begin{figure}[tbp]
  \centering
  \includegraphics[width=\linewidth]{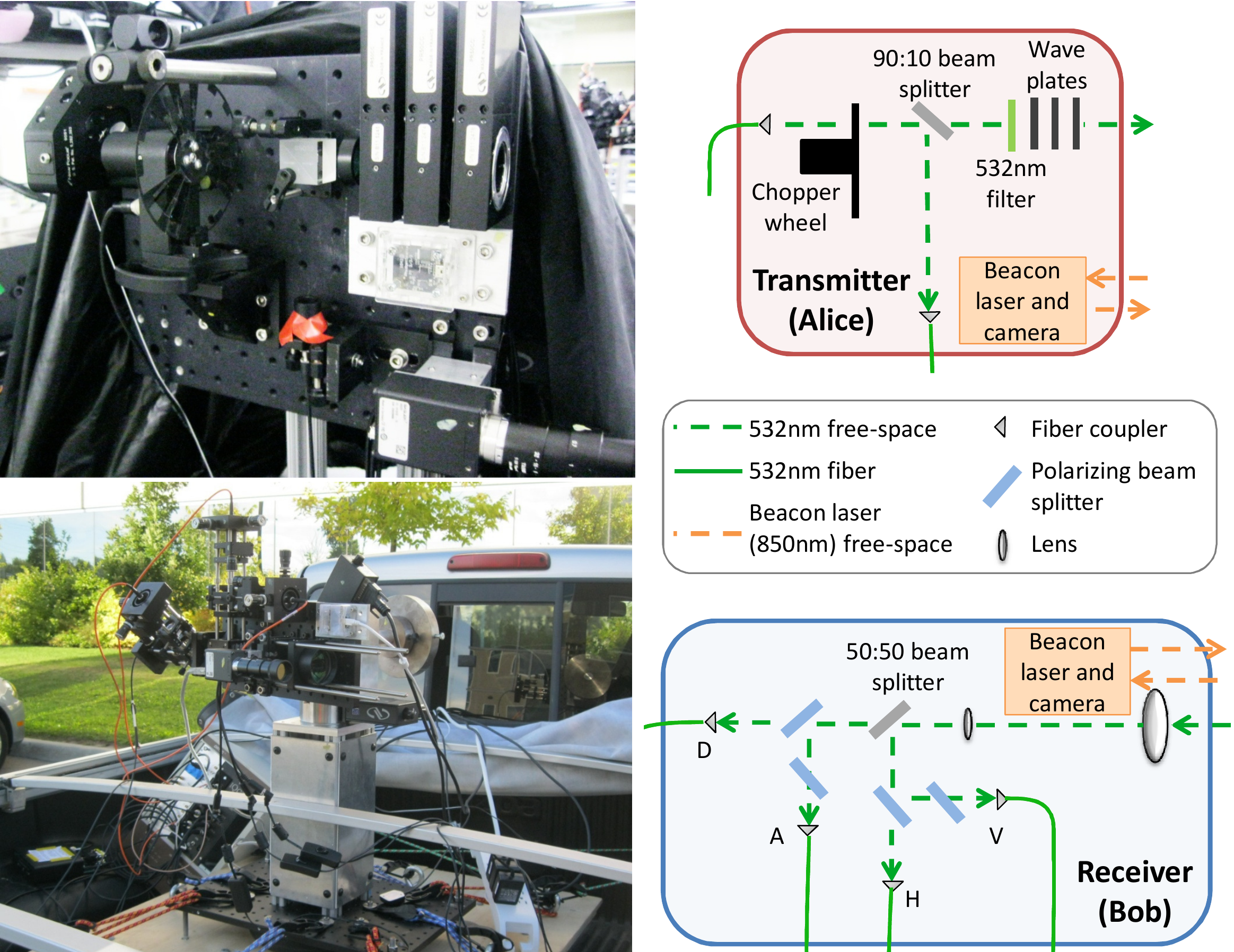}
  \caption[Transmitter and receiver]{Alice transmitter (top) and Bob receiver (bottom) apparatuses. The transmitter produces a $\SI{{\approx}10}{\mm}$ beam and includes a chopper wheel with polarizer films followed by a \SI{10}{\percent} reflective beam splitter with a fiber coupler in the reflected port, allowing us to tomographically characterize the polarization state of some of the transmitted photons. This characterization allows us to implement, in real time using a set of motorized wave plates, a compensation to the polarization drift of the states caused by the fiber to the transmitter. Only the signals that passed through the open slots of the chopper wheel are used for key generation. The receiver, which is mounted on the back of a pickup truck, implements a passive-basis-choice photon polarization measurement.}
  \label{fig.transmitter_receiver}
\end{figure}

The \SI{532}{\nm} pulses are sent to the transmitter (Fig.~\ref{fig.transmitter_receiver}, top), via an optic fiber, where they are collimated to a beam with ${\approx}\SI{10}{\mm}$ waist by a 1-inch diameter, \SI{30}{\mm} focal-length lens. The small beam waist was chosen to maintain simplicity of the transmitter (by not requiring large optics), and despite the additional loss caused by diffraction, the total loss remains significantly lower than that expected in a satellite uplink.

To correct for phases and rotations arising from the fiber transmission (which vary in time due to both temperature fluctuations in the building and mechanical stress on the fiber caused by the motion of the transmitter), the transmitter is equipped with a polarization characterization and compensation system. We use a modified optical chopper where upon each of six open slots has been placed a polarizer aligned to pass horizontal (H), vertical (V), diagonal (D), antidiagonal (A), right- (R), or left-circular (L) polarization. Some of the closed slots of the chopper wheel were removed to improve signal transmission. As a result, \SI{50}{\percent} of the signal pulses pass the chopper wheel unobstructed, \SI{20}{\percent} are polarized (either absorbed or passed, with probability dependent on the state overlap with the orientation of the polarizer) and the remaining \SI{30}{\percent} are blocked. A beam-splitter is used to send \SI{10}{\percent} of the passed signals to a multi-mode fiber which is coupled to a single-photon detector. A reference signal is given from the chopper wheel at each rotation to identify the position of the wheel (assuming a constant rotation speed between each rotation). The proportions of detected polarized signals are assessed each second and used to tomographically characterize the states at the transmitter. To return the measured states to the desired H, V, D, and A polarization states for QKD transmission, we optimize a theoretical model of a set of wave plates in rotation mounts, consisting of two quarter-wave plates on either side of a half-wave plate. We then apply the optimized rotations to actual wave plates mounted in motorized rotation stages.

With this, in contrast to other polarization alignment approaches~\cite{CWLWZ07, TTSTTFS11, ZZWYWHJCW14}, we are able to compensate, prior to the QKD signal leaving the telescope, \emph{any} phase or rotation which may be induced by the fiber, directly applying a unique compensation solution at all times. Our system can also compensate polarization drifts caused by other birefringent elements, such as glass and mirrors, as well as (slowly) moving elements such as those in a pointing mechanism, provided they are located before the polarization characterization. Moreover, given additional relative orientation information, our compensation optimizer could also easily correct for misalignment between receiver and transmitter frames. For our demonstration this was not implemented, as the receiver frame was inherently sufficiently aligned to the transmitter.

The receiver (Fig.~\ref{fig.transmitter_receiver}, bottom) consists of a 2-inch diameter input lens (\SI{100}{\mm} focal length) followed by a \SI{6.5}{\mm} diameter collimating lens (\SI{11}{\mm} focal length). A beam-splitter is used to implement a passive basis choice between the H/V basis (reflected port) and the D/A basis (transmitted port). The polarization measurement is performed using \SI{10}{\mm} polarizing beam-splitter cubes (PBS). An extra PBS (rotated \SI{90}{\degree} around the beam path) is added at the reflected port of each measurement PBS to suppress noise. In each of the four measurement outputs, the light is focused into multi-mode fibers (\SI{105}{\micro\meter} core diameter) using a 1-inch diameter, \SI{60}{\mm} focal-length collection lens and a \SI{12.5}{\mm} diameter, \SI{100}{\mm} focal-length focusing lens. The four fibers are connected to silicon avalanche photodiodes which detect the photons. The field of view of the receiver is \SI{0.02}{\degree}. Given some work to integrate the optical elements in a more compact form to minimize mass and volume, the simplicity of the overarching optical design of the analyzer makes it appropriate for implementation on a satellite platform. For our tests, the receiver platform is suspended on the truck with a small inner-tube and held in place with vertically sliding metal bars and elastic straps.

The transmitter and the receiver are each mounted on an orthogonally oriented pair of rotation stages, one with a horizontal (azimuth) plane of rotation and the other with a vertical (elevation) plane of rotation. On each motorized platform, a camera and three beacon lasers are included, allowing for the transmitter and receiver to each use the other's beacon signal deviation, measured on their camera, to control their motors. The pointing algorithm is based on velocity estimation of the beacon spot on the camera and its deviation from a nominal center, and adjusts the velocity of the motors to match the estimated velocity plus a proportional error term to reduce the deviation. Both of these corrections are based on a weighted average (exponentially decaying) of previous spot velocity estimates, in order to suppress high-frequency jitter.

\section{Results}

The size of the optical spot at the receiver had a measured diameter of about \SI{12}{\cm}. Following link acquisition, the total link loss at the receiver averaged \SI{30.6}{\dB}, mostly due to geometric effects, with other effects---including atmospheric transmittance, detector efficiency and losses in optical components---accounting for only \SI{7}{\dB} of loss. We calculate that the geometric effects contribute \SI{12}{\dB} loss from diffraction alone (due to the small \SI{10}{\mm} beam waist at the transmitter output), with an average of \SI{4.3}{\dB} of additional loss due to pointing error of the transmitter and atmospheric turbulence, and an average of \SI{7.3}{\dB} loss from the receiver pointing error. The receiver pointing error contributes more loss than the transmitter because road and motor-induced vibrations of the truck under motion result in increased jitter. These vibrations would be absent on a satellite platform.

\begin{figure}[tbp]
  \centering
  \includegraphics[width=0.7\linewidth]{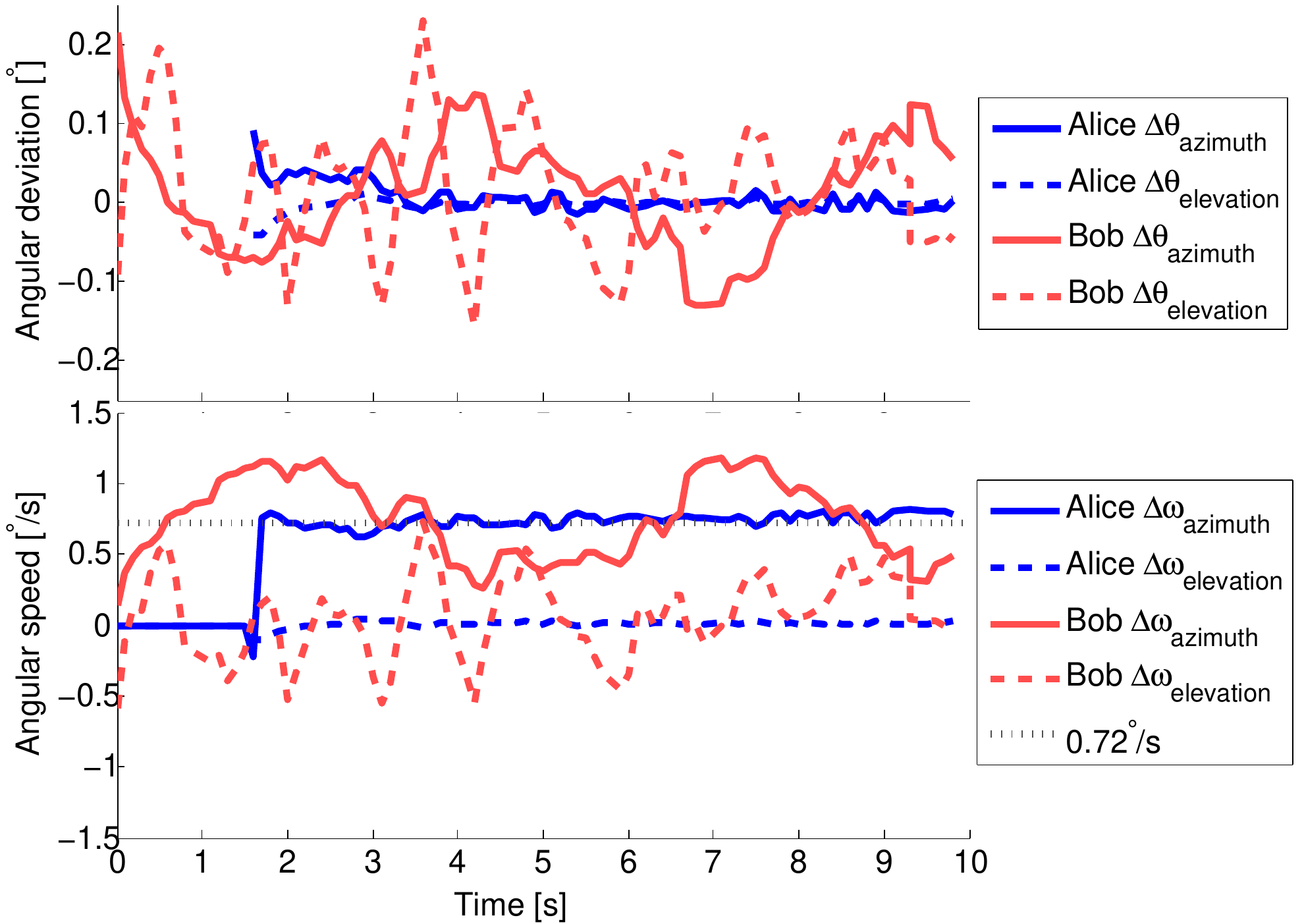}
  \caption[Angular deviation and speed]{Beacon angular deviation measured by the camera (top) and angular speed of the motors (bottom) during the \SI{33}{\km/\hour} test. The reported deviation on Alice (Bob) is the angular deviation of Bob's (Alice's) beacon as seen by Alice's (Bob's) camera. The increase in variation at the receiver (Bob) can be attributed to the instabilities produced by the motion of the truck and and curving motion of the truck. The angular speed at the transmitter (Alice) is more consistent, with only a small increase in the azimuthal axis (of ${\approx}\SI{0.03}{\degree/\s}$) during the test. The increased deviation in the azimuthal axis at the transmitter (Alice) is caused by the jitter in the mounting frame which can move more easily in the horizontal plane than the vertical plane. Before \SI{1.6}{\s}, Alice's motors are not moving because she has yet to acquire Bob's beacon signal. The gray dotted line represents the maximum angular speed of a \SI{600}{\km} LEO satellite (\SI{0.72}{\degree/\s}). }
  \label{fig.Pointing_offset_40km_h}
\end{figure}

During the experiment the truck is driven at a speed of \SI{33}{\km/h}, corresponding to an angular speed of \SI{0.75}{\degree/\s}, above the maximum angular speed of a LEO satellite platform at \SI{600}{\km} altitude: \SI{0.72}{\degree/\s} at zenith. The angular deviation of the motor positions, as measured by the imaged beacon spot on the camera, is shown in Fig.~\ref{fig.Pointing_offset_40km_h} (top). Initial acquisition and stabilization required approximately \SI{4}{\s} following first acquisition of the beacon. The mean deviation after stabilization of the link was measured to be \SI{0.005}{\degree} at the transmitter and \SI{0.06}{\degree} at the receiver. The extra deviation at the receiver is due to the jitter from the vibrations of the truck. (A pointing test with the truck stationary showed similar mean deviation to the measured transmitter deviation, \SI{0.005}{\degree}, at both parties.) Figure~\ref{fig.Pointing_offset_40km_h} (bottom) shows the angular velocity of the motors during the test. Both the transmitter and the receiver averaged an azimuthal angular speed above \SI{0.72}{\degree/\s}, with the transmitter (Alice) showing a more consistent angular speed.

\begin{figure}[b!]
  \centering
  \includegraphics[width=0.7\linewidth]{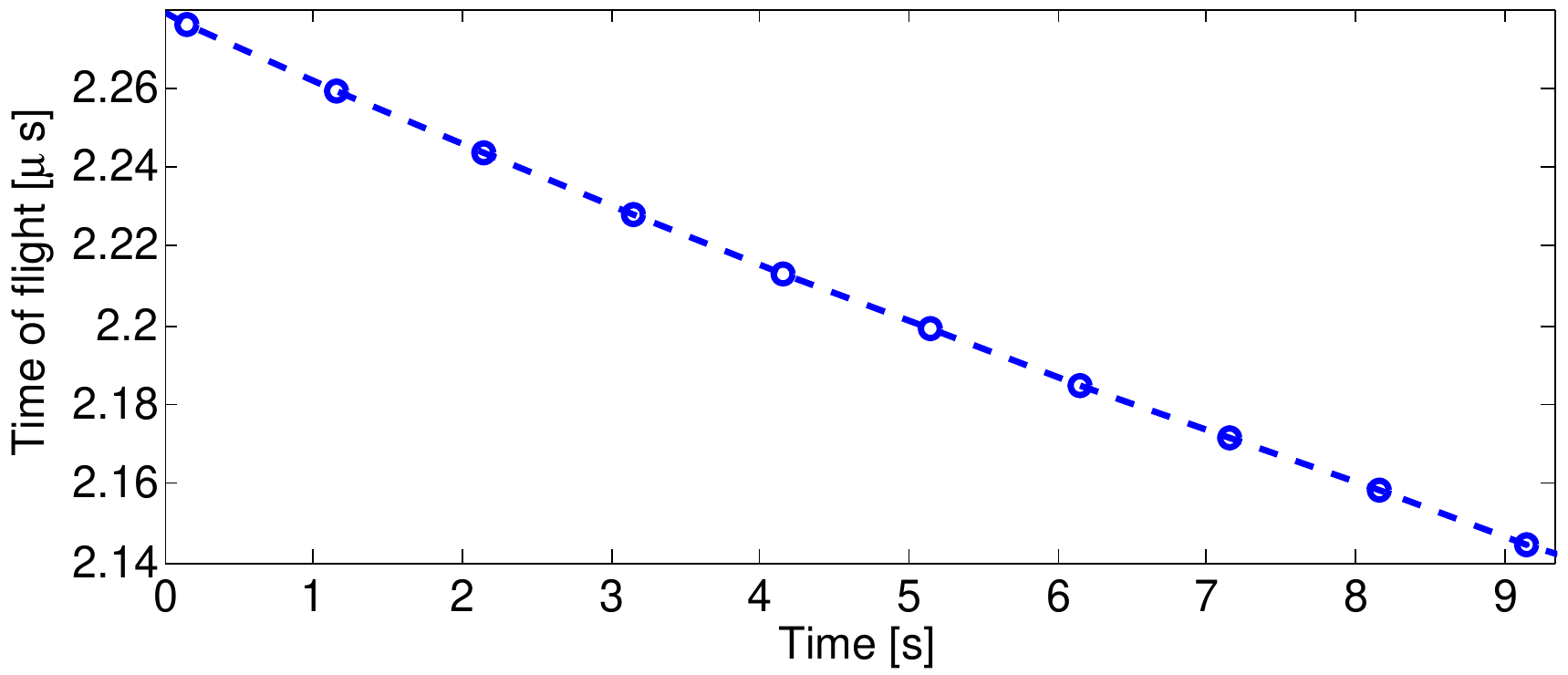}
  \caption[Time of flight from the transmitter to the receiver based on the GPS coordinates]{Time of flight from the transmitter to the receiver calculated from GPS coordinates. The time of flight changes almost linearly and can thus be effectively compensated using a first-order (linear) time-of-flight correction.}
  \label{fig.time-of-flight}
\end{figure}

The time of flight (Fig.~\ref{fig.time-of-flight}) is calculated from per-second GPS coordinates recorded during the experiment, and is subtracted from detection event times assuming a linear time-of-flight variation and extrapolating the position of the truck from the most recent per-second data using the velocity, also given by the GPS receiver. We observe a time-of-flight variation of around \SI{15}{\ns/\s}. A satellite time-of-flight variation, which could be as high as \SI{20}{\ms/\s} near the horizon (approaching zero at zenith, where the count rates are highest), could be corrected similarly, but may require greater fidelity, e.g. a higher-order interpolation incorporating predicted orbital positions.

\begin{figure}[tbp]
  \centering
  \includegraphics[width=0.7\linewidth]{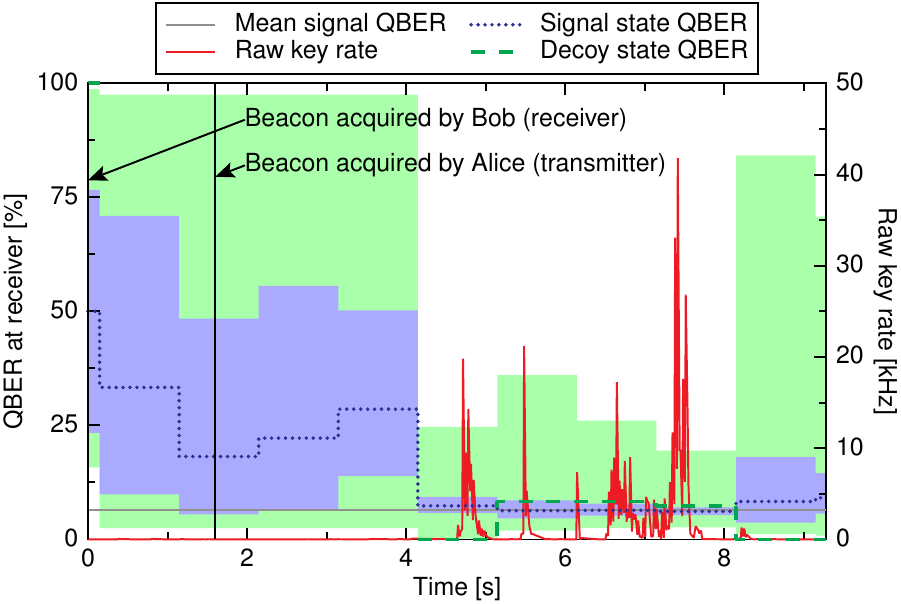}
  \caption[QBER and count rate measured at the receiver]{QBER and count rate measured at the receiver. The horizontal gray line represents the mean signal QBER of the data selected for the QKD protocol (\SI{6.55}{\percent}). The QBER drops when the count rate increases at \SIrange{4}{8}{\s}. The shaded regions (signal, blue; decoy, green) around the QBERs correspond to \SI{95}{\percent} central credible intervals. The large range of the credible interval of the decoy QBER is due to the low number of measured decoy states, with results near the beginning absent as no decoy states were measured at those times. The measured values are taken with a \SI{0.16}{\ns} coincidence window, with the QBER measured on a per-second basis and the counts measured on a \SI{2}{\ms} basis.}
  \label{fig.QBER_and_counts_40km_h}
\end{figure}

Figure~\ref{fig.QBER_and_counts_40km_h} shows the signal quantum bit error ratio (QBER), the decoy QBER, and the count rate during the test. The count rate increases once the link stabilizes after initial acquisition (which takes $\SI{{\approx}4}{\s}$), briefly reaching over \SI{41}{\kHz}. Even after acquisition, there remains a large amount of fluctuation due to the low pointing accuracy of the receiver, which varies outside the \SI{0.02}{\degree} field of view of the receiver. Both the signal and decoy QBER decrease as the counts increase, with a minimum measured signal QBER of \SI{6.16}{\percent}.

\begin{figure}[tbp]
  \centering
  \includegraphics[width=0.7\linewidth]{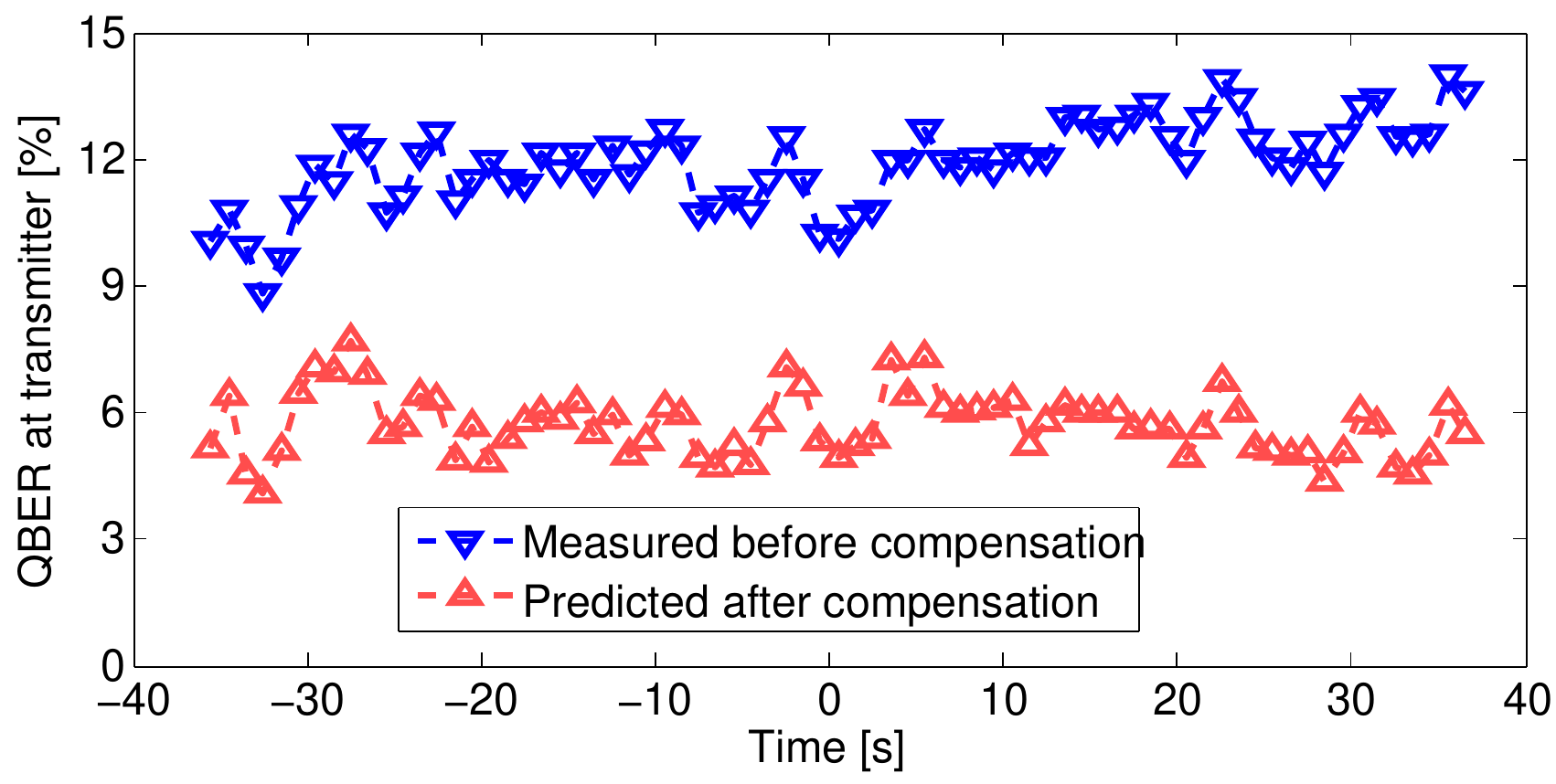}
  \caption[Measured pre-compensation and predicted post-compensation QBER at the transmitter]{Measured pre-compensation and predicted post-compensation QBER at the exit of Alice's transmitter. The time origin corresponds to the start of the optical signal acquisition. The polarization compensation system corrects the unitary induced by the fiber from the source to the transmitter and returns the polarization states to near their intrinsic QBER of ${\approx}\SI{6}{\percent}$.}
  \label{fig.Polnator_QBER_40km_h}
\end{figure}

Intrinsic QBER of the source is the primary contributor to the measured QBER. We indirectly measure the intrinsic QBER via the polarization compensation system, which measures the polarization states arriving at the transmitter and predicts those states after the optimized correction is applied by the wave plates. The measured QBER before compensation and the predicted QBER after compensation are shown in Fig.~\ref{fig.Polnator_QBER_40km_h}. Before compensation, the QBER varies around \SIrange{10}{13}{\percent}, increasing over the duration of the test, while the predicted QBER after compensation stays constant at around \SI{6}{\percent}. We conclude that the measured QBER at the receiver---\SIrange{6.5}{8}{\percent} when the count rate is above \num{1000}---is mostly caused by the intrinsic QBER found at the transmitter, with the additional \SIrange{0.5}{2}{\percent} owing to background noise at the receiver. While the pre-compensation QBER observed during the \SI{10}{\s} of data acquisition did not vary wildly, the polarization rotation induced by the motion of the optical fiber can be large and is effectively unpredictable. In addition, the observed long-term polarization drift was significant, necessitating optimization prior to each experimental attempt. These factors justify our usage of the real-time compensation system.

We examine the \SI{4}{\s} of data where the counts within each second total more than \num{1000}, effectively applying a signal-to-noise filter~\cite{EHMBLWJ12}---a filter with finer grain could potentially improve the results, given the high-frequency fluctuations of the count rate (Fig.~\ref{fig.QBER_and_counts_40km_h}). Post-processing steps, including error correction (using low-density parity check codes) and privacy amplification (based on reduced-Toeplitz-matrix two-universal hashes), are performed after the transmission, and secure key is extracted. The measured QKD parameters are reported in Table~\ref{tab.Moving_receiver_QKD_results}. (For details of our post-processing approach, see Ref.~\cite{BGHYMKLJ15}.) From these data, we extract \SI{160}{bits} of secure key, not considering finite-size statistical effects. A longer link duration would be required in order to extract secure key when including finite-size statistics---to extract secure key with finite-size statistics to ten standard deviations (a common heuristic~\cite{SLL09}), a total of \SI{13210}{\s} of data at the measured system performance would be required. 

\begin{table}[tbp]
\caption[Experimental QKD parameters of the moving receiver runs]{Experimentally measured QKD parameters of the moving receiver test. The parameters are based on a \SI{0.16}{\ns} coincidence window using data where the received counts exceeded \SI{1000}{\Hz}. The small window allows us to increase the signal-to-noise ratio (and thus reduce the QBER) at the cost of reduced raw key rate.}\label{tab.Moving_receiver_QKD_results}
\begin{center}
\begin{tabular}{cc} 
\hline\hline
\multicolumn{1}{>{\centering}p{5cm}}{\bf{Parameter}} & \multicolumn{1}{>{\centering}p{3cm}}{\bf{Value}} \\
\hline
Duration & \SI{4}{\s} \\
Signal average photon number & 0.495 \\
Decoy average photon number  &  0.120 \\
Signal QBER & \SI{6.55}{\percent} \\
Decoy QBER &  \SI{5.49}{\percent} \\
Signal gain & $5.86\times10^{-5}$ \\
Decoy gain & $1.5\times10^{-5}$ \\
Single photon gain lower bound & $3.72\times10^{-5}$ \\
Single photon QBER upper bound &  \SI{5.85}{\percent} \\
Vacuum yield & $1.35\times10^{-7}$ \\
Average loss &  \SI{30.6}{\dB} \\
Error correction efficiency & 1.15 \\
Raw key length & 11477~bits \\
Sifted key length & 5844~bits \\
Secure key length (asymptotic) & 160~bits \\
Secure key bit-string & 01001010100010101011 \\
& 00100110010001111010\\
& 00011100000000010000\\
& 10000000101111101111\\
& 01000001110100110101\\
& 10010011000001001001\\
& 10001100100001111111\\
& 00010111101111110111\\
\hline\hline
\end{tabular}
\end{center}
\end{table}

Improvement to the intrinsic QBER of the system could drastically reduce this link duration requirement. For example, in a similar system we recently demonstrated QKD at high loss, showing an asymptotic key rate of \SI{1.76}{\mega\bit/\s} at a loss of \SI{34.9}{\dB}~\cite{BGHYMKLJ15} (\SI{4.3}{\dB} above the measured \SI{30.6}{\dB} of our system), while QBERs of ${<}\SI{1}{\percent}$ have previously been achieved in other WCP QKD sources~\cite{TTSTTFS11}.

Following the completion of the experiment, modelling~\cite{JPBthesis14} showed that this QBER is largely due to reduced purity, efficiency imbalance between the two SFG crystals, and deviations in the modulator phase for two polarization states. This analysis was performed using measured state data obtained from the polarization compensation system, from which we calculate the purity and fidelity of the states at the transmitter. The mean purity and fidelity of the states was found to be \SI{91}{\percent} and \SI{94}{\percent} respectively. Our modeling shows a significant imbalance in the SFG efficiency, with one crystal showing a normalized efficiency of \SI{70}{\percent} compared to \SI{30}{\percent} in the other crystal, a difference of more than a factor 2. In addition, the modulator phases for two states (vertical and diagonal) show notable deviations from the desired values, explaining the lower fidelities measured in these two states when compared to the other two states. 

Another important effect that increased the intrinsic QBER is the reduced purity of the state, which may have been caused by unstable phase modulators, or unstable voltages applied to the modulators, in one polarity---the states with lower purity, vertical and diagonal, correspond to different applied voltage polarities than the states with higher purity, H and A~\cite{YMBHGMHJ13}. Other possible causes are spreading of the polarization components due to birefringence (given the short temporal width of the pulses), frequency-hopping of the \SI{1550}{\nm} laser, and noise due to background light. Accounting for this reduced purity, our modeled states predict an intrinsic QBER of \SI{5.33}{\percent}, comparable to the \SI{5.35}{\percent} predicted by the polarization compensation system.

The asymptotic-limit result of our experiment is sufficient to show the viability of QKD to a moving receiver platform travelling at an angular speed of up to \SI{0.75}{\degree/\s}. Of course, the useful optical contact time to a satellite (within a single pass) is limited to only a few hundred seconds. However, given sufficiently precise tracking, key extraction while accounting for finite-size fluctuation is nevertheless feasible under this limited time~\cite{BMHHEHKHDGLJ13, BGHYMKLJ15}.

\section{Conclusion}

We have thus demonstrated the feasibility of using a moving receiver platform for QKD. We constructed a custom pointing system, and used it to maintain a free-space quantum link over which BB84 decoy-state quantum signals were exchanged to a truck traveling at angular speeds consistent with a LEO satellite. Following timing analysis and post-processing steps, this was sufficient to generate asymptotically secure key. With reasonable improvements to the system, such as reducing the source intrinsic QBER, and the addition of a second-stage fine-pointing mechanism (both of which are largely engineering challenges with demonstrated solutions) we expect to achieve appreciable secure key bit rates with the inclusion of finite-size statistics. Such improvements will also allow us to demonstrate QKD to a moving receiver over longer distances and higher losses, completing necessary steps towards achieving long-distance satellite-mediated QKD networks.

\section*{Acknowledgments}

The authors would like to thank Christian Barna, Alexander Chuchin and Jennifer Fernick for assistance during the experimental tests. Support for this work by CFI, CIFAR, CryptoWorks21, CSA, FedDev Ontario, Industry Canada, Ontario Research Fund and NSERC is gratefully acknowledged. S.K. acknowledges support from Mike \& Ophelia Lazaridis Fellowship.

\end{document}